\documentclass[conference]{IEEEtran}
\IEEEoverridecommandlockouts
\usepackage{cite}
\usepackage{amsmath,amssymb,amsfonts}
\usepackage{algorithmic}
\usepackage{graphicx}
\usepackage{textcomp}
\usepackage{xcolor}
\usepackage{pdfpages}
\usepackage{subfigure}
\usepackage{booktabs}
\usepackage[justification=centering]{caption}
\usepackage{hyperref}
\hypersetup{
    colorlinks=true,
    linkcolor=blue,
    filecolor=magenta,
    urlcolor=cyan,
}

\def\BibTeX{{\rm B\kern-.05em{\sc i\kern-.025em b}\kern-.08em
    T\kern-.1667em\lower.7ex\hbox{E}\kern-.125emX}}
\begin{document}

\title{Target Speech Extraction with Pre-trained AV-HuBERT and Mask-And-Recover Strategy\\

 }

\makeatletter
\newcommand{\linebreakand}{%
  \end{@IEEEauthorhalign}
  \hfill\mbox{}\par
  \mbox{}\hfill\begin{@IEEEauthorhalign}
}
\makeatother

\author{\IEEEauthorblockN{1\textsuperscript{st} Wenxuan Wu}
\IEEEauthorblockA{\textit{The Chinese University of Hong Kong} \\ 
Hong Kong SAR, China \\
wwu@se.cuhk.edu.hk}
\and
\IEEEauthorblockN{2\textsuperscript{nd} Xueyuan Chen}
\IEEEauthorblockA{\textit{The Chinese University of Hong Kong} \\ 
Hong Kong SAR, China \\
xychen@se.cuhk.edu.hk}
\and
\IEEEauthorblockN{3\textsuperscript{rd} Xixin Wu}
\IEEEauthorblockA{\textit{The Chinese University of Hong Kong} \\ 
Hong Kong SAR, China \\
wuxx@se.cuhk.edu.hk}
\linebreakand 
\IEEEauthorblockN{4\textsuperscript{th} Haizhou Li}
\IEEEauthorblockA{\textit{The Chinese University of Hong Kong} \\ 
Shenzhen, China \\
haizhouli@cuhk.edu.cn}
\and
\IEEEauthorblockN{5\textsuperscript{th} Helen Meng}
\IEEEauthorblockA{\textit{The Chinese University of Hong Kong} \\ 
Hong Kong SAR, China \\
hmmeng@se.cuhk.edu.hk}

}
\maketitle
\vspace{-10pt}
\begin{abstract}
Audio-visual target speech extraction (AV-TSE) is one of the enabling technologies in robotics and many audio-visual applications. One of the challenges of AV-TSE is how to effectively utilize audio-visual synchronization information in the process. AV-HuBERT can be a useful pre-trained model for lip-reading, which has not been adopted by AV-TSE. In this paper, we would like to explore the way to integrate a pre-trained AV-HuBERT into our AV-TSE system. We have good reasons to expect an improved performance. To benefit from the inter and intra-modality correlations, we also propose a novel Mask-And-Recover (MAR) strategy for self-supervised learning. The experimental results on the VoxCeleb2 dataset show that our proposed model outperforms the baselines both in terms of subjective and objective metrics, suggesting that the pre-trained AV-HuBERT model provides more informative visual cues for target speech extraction.
Furthermore, through a comparative study, we confirm that the proposed Mask-And-Recover strategy is significantly effective. 
\end{abstract}

\vspace{-10pt}
\section{Introduction}
\label{sec:intro}

The quest for the understanding of human's cocktail party effect and the implementation of automatic speech separation has never stopped. On one hand, a speech separation system is generally seen as the front-end for some downstream tasks such as speech recognition\cite{token2vec}, speaker verification\cite{spkverify}, speech translation\cite{speech-translation}, speech synthesis\cite{chen2022hilvoice}, etc. On the other hand, speech separation technology is often inspired by findings in auditory cognition of the human brain\cite{brain}. \\
\indent With the advent of deep learning, effective speech separation models are proposed, such as uPIT\cite{upit}, and Conv-TasNet\cite{Conv-TasNet}, which shown good performance. However, as the number of speakers in a cocktail mixture is often required in advance for speech separation to work, this limits the scope of applications. Furthermore, a typical speech separation system seeks to disentangle multiple speakers from a cocktail mixture, speech separation sometimes suffers from speaker permutation problem~\cite{upit}. The target speaker extraction (TSE) seeks to extract target speaker's voice from a mixture speech conditioned on certain cues\cite{TSE-review}. Such a task is more consistent with the human selective auditory attention process in a cocktail party~\cite{select-listen}.

\indent For effective TSE, the choice of proper speaker cues remains a topic of study. There have been studies    
to utilize a pre-recorded speech sample from the target speaker as such speaker cue \cite{SpeakerBeam}. More recent studies utilize the speaker embedding extracted from the pre-recorded speech as the speaker cues that have improved the performance \cite{VoiceFilter}\cite{SpEx}\cite{SpEx+}.
However, such techniques suffer from some mismatch between the provided speech cue and the target speech in the mixture, such as different recording scenarios and different speech semantics from the same speaker, which is also called the intra-speaker variety problem in the TSE system\cite{TSE-review}.  To tackle such intra-speaker variety, more stable and noise-invariant visual cues are expected.
 
Among all the visual cues, lip movement could be recognized as the most informative cue\cite{Rethinking}. Nonetheless, visual occlusion problems may still occur in practical application scenarios. When the camera moves or the target speaker turns around, the target speaker's frontal view changes. To deal with such challenges, ImagineNET\cite{ImagineNET} utilized some extra visual refiner blocks and visual decoders to overcome the missing visual cues. Considering the noise and reverberation invariant properties as well as the advanced solutions towards visual occlusion, lip movements remain the most effective target speaker cue for the TSE systems. \\\indent
In terms of network architecture, there are generally two types, namely time-frequency domain and time domain\cite{TSE-review} systems. 
VisualVoice is a typical time-frequency domain system and utilizes the U-Net architecture as the backbone of the separation module \cite{VisualVoice}. Such time-frequency domain TSE systems typically suffer from performance issues such as imperfect phase recovery. 
Most current TSE systems adopted time-domain methods such as Conv-TasNet\cite{Conv-TasNet}. Conv-TasNet utilized the original mixture speech signal as input and passed it to the convolution layers to get the speech embeddings\cite{Conv-TasNet}. Such convolution filters are designed to simulate the human auditory system and the obtained speech embeddings approximate spectrogram in the time-frequency domain. In this paper, we adopt a time-domain backbone architecture. 
\\\indent Currently, speech foundation models have shown dominant performance in many speech tasks such as speech recognition\cite{ssl_speech_review} and speech synthesis\cite{chen2022character}\cite{chen2022unsupervised}. Thanks to the self-supervised learning strategy and a plentiful of unlabeled speech data,  noise robust speech representations could be obtained from speech foundation models than just using convolution blocks in a supervised way\cite{chen2023stylespeech}. Although most speech foundation models are pre-trained on speech recognition tasks such as wav2vec 2.0\cite{wav2vec2} and HuBERT\cite{HuBERT}, recent studies have also proven the feasibility of using the speech foundation models for speech separation tasks. In \cite{Huang2022InvestigatingSL}, the author investigated the performances of 13 different speech foundation models on speech separation tasks and verified the robustness of the self-supervised speech features for such waveform generation tasks. To obtain better performance, more recent work concatenated both spectrogram and speech embeddings obtained by WavLM\cite{WavLM} as input for downstream separation modules, which also achieved significant improvements with scaled-up training data\cite{wavlm-ss}. Despite some exploration of speech foundation models for speech separation tasks, speech foundation model has not been utilized in the current TSE system.\\\indent
Recently, AV-HuBERT achieved great success in lip-reading tasks\cite{chen2024exploiting}, showing its strong ability of capturing audio-visual synchronization\cite{AVhubert}. To benefit from such robust audio-visual synchronization knowledge, we integrate pre-trained AV-HuBERT layers into our TSE system. Furthermore, to facilitate the alignment between audio feature space and visual feature space, a novel Mask-And-Recovery (MAR) strategy has been applied to our TSE system.  With the integrated AV-HuBERT layers and additional MAR strategy, we propose the AVHuMAR-TSE system.
The contributions of this paper could be summarized in three folds:
\begin{itemize}
\item  First, we integrate the pre-trained AV-HuBERT layers into our proposed audio-visual TSE system, which is called the AVHuBERT-TSE system. To the best of our knowledge, this is the first attempt to combine the  AV-TSE system with the audio-visual foundation model. \\
\item  To enhance both intra and inter-modality alignments, we further propose the AVHuMAR-TSE system, which jointly optimizes the pre-trained AVHuBERT-TSE system and the integrated MAR block. Experimental improvements demonstrate the effectiveness of the proposed Mask-And-Recovery (MAR) strategy.\\
\item  To verify the effectiveness of the proposed MAR strategy, we experiment with different mask durations for the mixture speech and find the best configuration for our AVHuMAR-TSE system.  
\end{itemize}
\section{Method}

\subsection{AVHuBERT-TSE System}
\begin{figure*}[t]
\centering
 \includegraphics[scale=0.5]{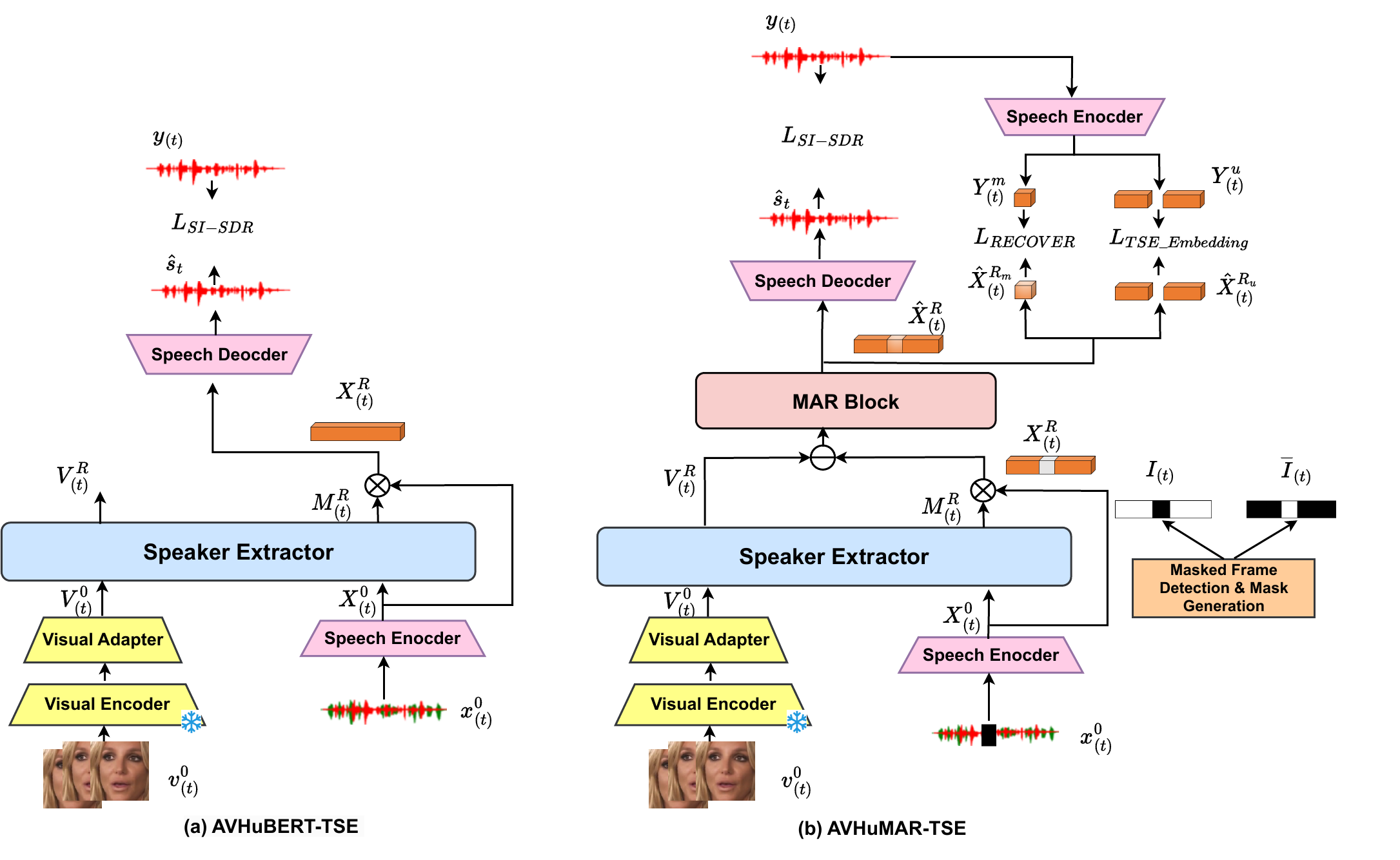}
\caption{The overall architecture of the proposed AVHuMAR-TSE system. The system (a) on the left is the basic AVHuMAR-TSE system without MAR block, which is called the AVHuBERT-TSE system. The system (b) on the right is the complete AVHuMAR-TSE system.}
\label{Figure1}
\end{figure*}
\subsubsection{Overview of the AVHuBERT-TSE System}
 AVHuBERT-TSE system followed the general framework of the mask-based time domain TSE system, which consists of three parts including the speech encoder, speech decoder, and speaker extractor, To obtain iteratively refined visual cues, the cue encoder will be integrated into the the speaker extractor in our AVHuBERT-TSE system.
 
 More specifically, as shown in Fig. \ref{Figure1}, the speech encoder takes mixture speech signal $x^0_{(t)}$ as input and outputs the mixture speech embedding $X^0_{(t)}$. The visual encoder takes corresponding visual frames $v^0_{(t)}$ as input and outputs target speaker lip embedding $V^0_{(t)}$. Note that the visual encoder follows the structure in MuSE \cite{muse}. The visual encoder is pre-trained on the visual speech recognition task and weights are frozen in our AVHuBERT-TSE system. The obtained target speaker lip embedding $V^0_{(t)}$ is supposed to encode the viseme-phoneme correspondence.
  Similar to \cite{tdse} \cite{select-listen} \cite{usev}, a visual adapter is followed after the visual encoder to obtain refined visual cue embedding $V^0_{(t)}$. The visual adapter follows the structure in \textit{reentry} \cite{select-listen}.
 As for the speaker extractor, it perceives both $V^0_{(t)}$ and $X^0_{(t)}$  to estimate the final target speaker's speech mask $M^R_{(t)}$.  Note that the final refined visual cue $V^R_{(t)}$ will also be output for further use in a more advanced AVHuMAR-TSE system. The final estimated mask $M^R_{(t)}$  will then element-wise multiplied with $X^0_{(t)}$ to get the final estimated target speech embedding $X^R{(t)}$. The speech decoder takes $X^R_{(t)}$ as input and reconstructs the final extracted target speech $\hat{s}_{(t)}$. Note that all the speech encoders and speech decoders of the AVHuBERT-TSE system keep the same structure and share weights during training.
\begin{figure}[t]
\centering
 \centering
 \includegraphics[scale=0.5]{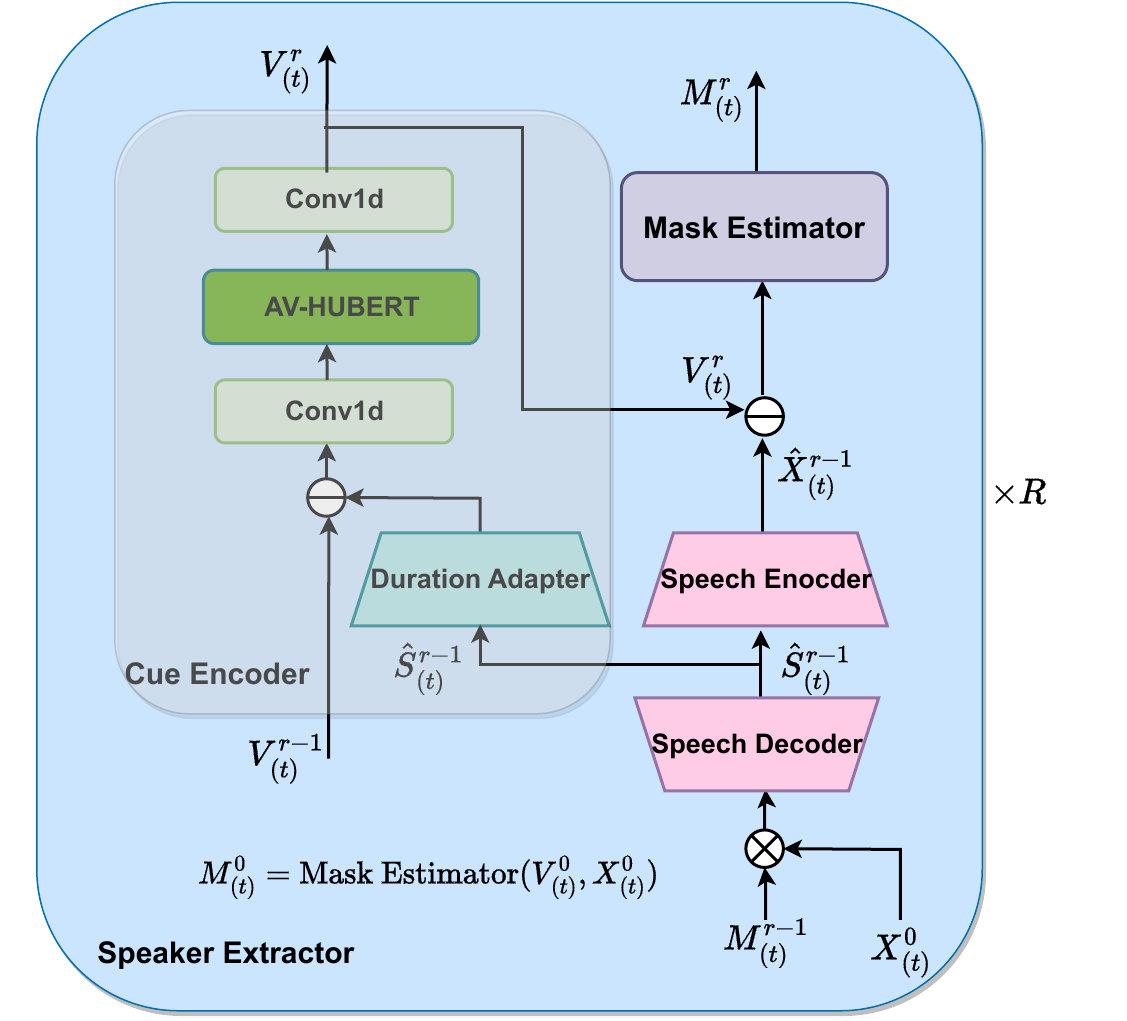}
\caption{The Speaker Extractor will be repeated R times. For $r-th$ Speaker Extractor, it mainly contains a Cue Encoder to refine the target speaker's visual cue $V^r_{(t)}$ and a Mask Estimator to predict the target speech mask $M^r_{(t)}$. The Speech Decoder and Speech Encoder are utilized to reconstruct and encode the intermediate estimated speech $\hat{S}^r_{(t)}$, respectively. Note that the initial target speech mask $M^0_{(t)}$ will be predicted conditioned on the initial visual cue $V^0_{(t)}$ and mixture speech embedding $X^0_{(t)}$.}
\label{Figure2}
\end{figure}
 \subsubsection{Speaker Extractor}
The speaker extractor aims at estimating the target speech mask from the mixture speech embedding $X^0_{(t)}$ guided by the visual cue. As shown in Fig. \ref{Figure2}, the speaker extractor mainly consists of two parts, i.e. mask estimator and cue encoder. Since the iterative mask estimators have been shown effective for TSE systems \cite{muse}\cite{tdse}, AVHuBERT-TSE also follows this structure. More specifically, the speaker extractor takes $X^0_{(t)}$ and $V^0_{(t)}$ as initial input. Then  $X^0_{(t)}$ and $V^0_{(t)}$ will be concatenated on channel dimension. A coarse target speech mask $M^0_{(t)}$ will be predicted by the first mask estimator conditioned on the concatenated audio-visual embedding. For simplicity and consistency, this initial mask estimating process is not drawn in Fig. \ref{Figure2} but described with $M^{0}_{(t)}$ = MASK ESTIMATOR$(V^{0}_{(t)}, X^{0}_{(t)})$. More generally, each speaker extractor receives both $M^{r-1}_{(t)}$ and $V^{r-1}_{(t)}$ from the last speaker extractor. The intermediate estimated speech $S^{r-1}_{(t)}$ will be first predicted by the element-wise multiplication of $M^{r-1}_{(t)}$ and $X^0_{(t)}$ and then generated through speech decoder. \\\indent The intermediate estimated speech $S^{r-1}_{(t)}$ will be utilized for two branches. For the first branch, $S^{r-1}_{(t)}$  will be passed to the cue encoder to obtain the refined visual cue $V^r_{(t)}$. Considering of computing efficiency, $S^{r-1}_{(t)}$ will be passed through five Conv1d layers called duration adapter to align with $V^{r-1}_{(t)}$ temporally. The duration adapter structure is similar to the HuBERT front-end while the kernel sizes for Conv1d layers are changed to (2,2,2,2,2). After alignment, $V^{r-1}_{(t)}$ and $S^{r-1}_{(t)}$ are concatenated temporally and passed into a Conv1d layer before input to the AV-HuBERT layers. Such a design aims at mitigating the adverse effect caused by front-end mismatch with the original AV-HuBERT. Besides, it will convert the channel dimension of the audio-visual cue embedding to adapt the transformer layer channel dimension. Previous works have shown that for the self-supervised speech foundation models, the acoustic features are learned in the shallow layers while high-level linguistic features are encoded by the deeper layers \cite{adapter}\cite{sidecar}. Motivated by this conclusion, AVHuBERT-TSE only adapted the first four layers of pre-trained AV-HuBERT to obtain refined visual cues. Another symmetric Conv1d layer is followed by the AV-HuBERT layers to transfer the cue embedding channel back. Note that the two additional symmetric Conv1d layers could be helpful to mitigate the  “transplant rejection” problem found in \cite{sidecar}. Besides, the visual cue embedding $V^{r}_{(t)}$  refined by the cue encoder is expected to contain rich audio-visual correspondence for the target speaker and have a similar resolution with the intermediate estimated speech embedding $X^{r-1}_{(t)}$. \\
\indent As for the second branch of intermediate estimated speech $S^{r-1}_{(t)}$, it will be passed through the speech encoder to obtain the intermediate estimated speech embedding $X^{r-1}_{(t)}$. The intermediate estimated speech embedding $X^{r-1}_{(t)}$  is then concatenated with the refined visual cue $V^{r}_{(t)}$ and passed into the mask estimator to predict the refined target speech mask $M^{r}_{(t)}$.\\
\indent Note that the speaker extractor will be repeated R times to learn hierarchical target speech features. To mitigate the adverse effect caused by model intrinsic uncertainty, all R speaker extractors will share the same weights during training.

 

\subsection{AVHuMAR-TSE System}
\subsubsection{Mask-And-Recovery Strategy}
 The extraction performance of the AV-TSE system guided by lip movement severely relies on the viseme-phoneme mapping correspondence in the feature space as well as the audio-visual temporal synchronization \cite{still}\cite{still-image}. 
 As a result, visual cues have been hierarchically refined in our AVHuBERT-TSE system to enhance robustness. More specifically, for the visual front-end, certain prior knowledge about viseme-phoneme mapping has been learned during the visual speech recognition pre-task. The visual cues obtained from such a visual front-end will be further adapted by the powerful lip-reading expert in our proposed cue encoder. However, since the viseme-phoneme mapping is not bijective\cite{speech-inpainting}\cite{Phoneme-to-viseme}, such an AV-TSE system is imperfect. Although the lip movement cue could be noise and reverberation invariant, the model may still be confused when different speakers have similar lip movements in the same mixture frame. Besides, the continuous intermediate estimated target speech frames contain rich context information, which has not been fully utilized as extraction guidance in the previous TSE systems. To fully use such intra-modality correlation in speech contexts and alleviate the extraction performance degradation caused by the viseme-phoneme mapping ambiguity, a novel Mask-And-Recover strategy has been applied to the proposed AVHuMAR-TSE system. 
\subsubsection{Overview of the AVHuMAR-TSE System}
 The AVHuMAR-TSE system is modified based on the AVHuBERT-TSE system. More specifically, as shown in Fig. \ref{Figure1} (b), compared to the AVHuBERT-TSE system shown in (a), the visual encoder, visual adapter, speech encoder, speech decoder, and speaker extractor will keep the same architecture while an additional Mask-And-Recover (MAR) Block will be inserted between the final speech decoder and the last speaker extractor. As the name described,  the MAR strategy will mask certain frames of the mixture speech waveform $x^{0}_{(t)}$  temporally. Then the masked mixture speech waveform $x^{0}_{(t)}$  together with the corresponding intact lip frames $v^{0}_{(t)}$ will be used as input for training AVHuMAR-TSE system. \\
\indent Note that the masked region of the mixture speech embedding $X^{0}_{(t)}$ may be temporally compressed by the speech encoder. For generalization consideration, an automatic masked frame detection step is necessary before passing masked $X^{0}_{(t)}$ into the speaker extractor. Since the speech encoder consists of a Conv1d layer followed by a Relu activation function, to avoid confusing the masked and non-masked region, such an automatic masked frame detection step will be conducted before the Relu operation. Note that the embedding value of the masked region will be the global extreme minimum after passing through Conv1d layers without bias, the detection step only needs to find the continuous indexes with global extreme minimum. To filter out those potential non-artificial small masked regions, further detection will be conducted with a proper threshold. According to the detected masked regions of $X^{0}_{(t)}$,  a mask $I_{(t)}$ and an inverse mask $\overline{I}_{(t)}$ with the same temporal shape with $X^{0}_{(t)}$ will be generated for loss computation.  
As shown in Fig. \ref{Figure1} (b), the masked region in target speech embedding $X^{R}_{(t)}$ keeps consistent with $X^{0}_{(t)}$. Instead of passing $X^{R}_{(t)}$ into the speech decoder directly, the $X^{R}_{(t)}$ will be concatenated with the last refined visual cue $V^{R}_{(t)}$ on channel dimension and then be passed into MAR Block.\\\indent The MAR Block plays two roles in the AVHuMAR-TSE system. First, it will be used to predict the masked target speech embedding. To predict the masked region, both unmasked target speech embedding region and refined visual cues will be utilized as guidance. The unmasked target speech embedding region could provide rich speech context information and push the model to learn intra-modality correlation. The refined visual cues could help the model to learn a direct mapping from the target speaker's lip movements to the masked speech embedding region. Second, the MAR Block will be jointly optimized with the AVHuBERT-TSE system shown in Fig. \ref{Figure1} (a), thereby enabling further refinement of the target speech embedding estimated by the AVHuBERT-TSE system. 
\\\indent The structure of the MAR Block simply consists of 4 transformer layers. To perform the embedding level loss, the ground truth of the target speech waveform $y_{(t)}$ will be input to a speech encoder to get its embedding $Y_{(t)}$. With the automatically detected mask $I_{(t)}$ and inverse mask $\overline{I}_{(t)}$, the masked region loss and the unmasked region loss will be computed, respectively. 
\subsection{Two-Stage Training Strategy}
The AVHuMAR-TSE system contains two training stages. In the first training stage, the intact mixture of speech and intact visual frames will be used for training. The system does not contain a MAR Block at this stage.  The overall architecture of the system has been shown in Fig. \ref{Figure1} (a). The goal of the first training stage is to obtain a well-trained AVHuBERT-TSE system. In the second training stage, the masked mixture speech and intact visual frames will be used for training, which is shown in Fig. \ref{Figure1} (b). In addition to the AVHuBERT-TSE system, an additional speech encoder and the MAR Block will be added in the second training stage. The additional speech encoder will share the same weights with other speech encoders. All the modules in the AVHuBERT-TSE system, including the additional speech encoder will be fine-tuned based on the first training stage, while only the MAR block will be trained from scratch.
\subsection{Loss Functions}
Both the first and second training stages will perform SI-SDR loss between the predicted target speech signal and the ground truth of the target speech signal, which is shown in \eqref{sisdr},
\begin{equation}
    L_{SI-SDR} (y_{t},\hat{s}_{t}) = -10\log_{10}(\frac{\frac{||<\hat{s}_{t}, y_{t}>y_{t}||^2}{ ||y_{t}||^2}} { ||\hat{s}_{t} - \frac{<\hat{s}_{t}, y_{t}>y_{t}}{ ||y_{t}||^2}||}).
    \label{sisdr}
\end{equation}
\indent During the second training stage, two extra mean squared error (MSE) \cite{MEAN-SQUARED-ERROR} loss functions will be added. To recover the masked embedding region, one of the MSE loss functions will be calculated for the masked region of $\hat{X}^{R}_{(t)}$, which is called $L_{RECOVER}$. To keep and refine the TSE extraction performance, another MSE loss will be calculated for the unmasked region of $\hat{X}^{R}_{(t)}$, which is called $L_{TSE\_Embedding}$. The calculation details for the MSE loss functions are described in \eqref{mse},
\begin{equation}
\begin{aligned}
    &L_{RECOVER}(\hat{X}^{R_m}_{(t)},Y^{m}_{(t)}) = {MSE}(\hat{X}^{R_m}_{(t)}, Y^{m}_{(t)}),\\
    &L_{TSE\_Embedding}(\hat{X}^{R_u}_{(t)}, Y^{u}_{(t)}) = {MSE}(\hat{X}^{R_u}_{(t)}, Y^{u}_{(t)}),\\
    \text{where,}\\
    &\hat{X}^{R_m}_{(t)} = \hat{X}^{R}_{(t)} \odot \overline{{I}}_{(t)},\\
    &\hat{X}^{R_u}_{(t)} = \hat{X}^{R}_{(t)} \odot {I}_{(t)},\\
   & Y^{m}_{(t)}  = Y_{(t)}  \odot \overline{{I}}_{(t)},\\
     &Y^{u}_{(t)}  = Y_{(t)}  \odot {I}_{(t)}.
  \end{aligned}
  \label{mse}
\end{equation}
\indent Specifically, $\hat{X}^{R_m}_{(t)}$ and $Y^{m}_{(t)}$ denote the masked region of predicted target speech embedding and corresponding ground truth embedding. $\hat{X}^{R_m}_{(t)} $ and $Y^{m}_{(t)}$  are obtained by the element-wise multiplication of the inverse mask $\overline{I}_{(t)}$ and $\hat{X}^{R}_{(t)} $, $Y_{(t)} $, respectively. On the contrary, $\hat{X}^{R_u}_{(t)} $ and $Y^{u}_{(t)}$ denote the unmasked region of predicted target speech embedding and unmasked region of corresponding ground truth embedding. $\hat{X}^{R_u}_{(t)} $ and $Y^{u}_{(t)}$ are also obtained with the similar  way. \\\indent The loss function for the AVHuMAR-TSE system in the second training stage contains three parts and is denoted in \eqref{finalloss}, 
\begin{equation}
\begin{aligned}
   &L(y_{t},\hat{s}_{t},\hat{X}^{R_m}_{(t)},Y^{m}_{(t)},\hat{X}^{R_u}_{(t)}, Y^{u}_{(t)})\\ = & \alpha * L_{SI-SDR} (y_{t},\hat{s}_{t})+ \beta* L_{RECOVER}(\hat{X}^{R_m}_{(t)},Y^{m}_{(t)})\\&+\gamma*L_{TSE\_Embedding}(\hat{X}^{R_u}_{(t)}, Y^{u}_{(t)}),
  \end{aligned}
  \label{finalloss}
\end{equation}
\indent  We use ${\alpha,\beta,\gamma }$ as scale factors to balance the three loss function parts, respectively. 
\section{Experimental setting}
\subsection{Dataset}
\begin{table*}[!ht]
\begin{center}
\caption{\centering AVHuMAR-TSE and Baseline Performances on Test Set.}
\label{tab:1}
\begin{tabular}{p{0.1\textwidth}p{0.18\textwidth}p{0.1\textwidth}p{0.08\textwidth}p{0.08\textwidth}p{0.08 \textwidth}p{0.08 \textwidth}p{0.08\textwidth}}  
\toprule
  \textbf{System} & \textbf{Model} &\textbf{Cue Type}&\textbf{SI-SDR $(\uparrow)$} & \textbf{SI-SDRi $(\uparrow)$} &\textbf{SDR $(\uparrow)$}& \textbf{PESQ $(\uparrow)$} &\textbf{STOI $(\uparrow)$}             \\ \toprule
 Baseline 1&AV-ConvTasNet\cite{tdse} &Lip& 10.725 & 10.771&11.099 & 2.592&0.859  \\
Baseline 2&USEV\cite{usev}& Lip&10.785 & 10.829&11.332 & 2.646&0.862  \\
  Baseline 3&MuSE\cite{muse} & Lip + Speaker&11.458 & 11.506&11.836 & 2.706&0.873 \\
\toprule
Proposed 1&AVHuMAR-TSE(w/o MAR)& Lip & 11.728 & 11.771&12.043 & 2.765  &0.878\\
Proposed 2&AVHuMAR-TSE & Lip &\textbf{12.331} & \textbf{12.379}&\textbf{12.726} & \textbf{2.922}&\textbf{0.887} \\
\bottomrule
\end{tabular}
\end{center}
\end{table*}
To evaluate the performance of the AVHuMAR-TSE system, we simulate 2-speaker mixture dataset from the VoxCeleb2\cite{voxceleb2} dataset. Similar to \cite{muse}, 48,000 utterances from 800 speakers are selected for the training set and 36,237 utterances from 118 speakers are selected for the test set. More specifically, we simulate 20000, 5000, and 3000 utterances for the training set, validation set, and test set,  respectively. Each target speech utterance has been mixed with the interfering speech utterance at a random Signal-to-Noise ratio (SNR) between -10 dB and 10 dB. The speech sampling rate is 16000 and the video frame is 25 FPS. Note that compared to the training set and validation set, all speakers in the test set are unseen speakers. Additionally, all the utterances are clipped to 4 seconds during training and 4-6 seconds during inference. For the second training stage, a random segment of each utterance in the training set will be masked with zero value. To find the optimal mask duration of the proposed AVHuMAR-TSE system, we simulate the mask duration gap equal to 100ms, 200ms, 300ms, 400ms, 500ms, and 600ms, respectively. Each mask duration gap is applied to the entire training set.
\subsection{Baseline and Evaluation Metric}
Since the AVHuMAR-TSE system is a time-domain audio-visual TSE system. To make a fair comparison, we also select three time-domain AV-TSE systems including AV-ConvTasNet\cite{tdse}, USEV\cite{usev}, and MuSE\cite{muse} as our baseline systems. \\
\indent For the evaluation metrics, we select the scale-invariant signal-to-noise ratio (SI-SDR)\cite{SDR}, the SI-SDR improvement (SI-SDRi), and the signal-to-noise ratio (SDR)\cite{SDR} as subjective metrics, which are normally used to evaluate the separation quality of extracted target speech. To evaluate the speech quality and intelligibility, we also use the perceptual evaluation of speech quality (PESQ)\cite{pesq} and the short term objective intelligibility (STOI) \cite{stoi} as objective metrics. The higher the better for all metrics.
 
\subsection{Implementation Details}
We re-implement three baseline systems with float32 precision. 
All the baseline results are reported on our test set. For the proposed AVHuMAR-TSE system, all speech encoders, speech decoders, and visual encoders follow the settings in \cite{muse}, 
while the mask estimator follows the structure from \cite{tdse}. 
The model parameters (N, L, B, H, P, X, R) are set to (256, 40, 256, 512, 3, 7, 
 4), respectively.
The threshold for detecting masked regions is set to 20 continuous waveform samples. The scale factors $(\alpha,\beta,\gamma)$ of the loss function are set to (1,5,1).
\\\indent For the first training stage, we train 150 epochs with a learning rate of 0.00015 and select the best checkpoint depending on the performance of the validation set. For the second training stage, we load the best pre-trained checkpoint from the first training stage and then train for another 30 epochs with the same learning rate. In the second training stage, we utilize the mixture speech with different mask duration gaps as input and select the checkpoint with the best validation performance as our final system. After the second training stage, we report the test set performance of our final AVHuMAR-TSE system. For GPU, all the individual experiments of the AVHuMAR-TSE system are trained on 3 V100 GPUs, and the batch size is set to 2.
\section{Experimental Results}
\subsection{Comparison with Different Baseline Systems}
 
As shown in TABLE \ref{tab:1}, we report test set results of three baseline systems and our proposed AVHuMAR-TSE system. Among all the systems, only MuSE uses both the target speaker's lip movements and speaker labels as cues while the others only use the target speaker's lip movements as the cue. Among all the baseline systems, MuSE is the best baseline system in terms of both subjective and objective metrics, which could achieve 11.458 on SI-SDR, while AV-ConvTasNet and USEV could only achieve 10.725 and 10.785 respectively. The credits are given to the iterative mask estimators and additional target speaker labels. Compared to the baseline results, our proposed AVHuMAR-TSE could achieve the best performance on all the metrics with 12.331 on SI-SDR, 12.726 on SDR, 2.922 on PESQ, and 0.887 on STOI.  Such results are significantly higher than MuSE, and demonstrate the effectiveness of the AV-HuBERT cue encoder and the proposed MAR strategy. \\
\indent To explore the improvements brought by each module, we also report the test set performance of AVHuMAR-TSE without MAR blocks. As shown in TABLE \ref{tab:1},  the system could still achieve 12.043 on SDR, 2.765 on PESQ, and 0.878 on STOI, respectively. The performances on both subjective and objective metrics are better than MuSE. Note that MuSE utilizes the same visual cue for all mask estimators and still needs additional target speaker labels as input. The proposed AVHuMAR-TSE utilizes the pre-trained AV-HuBERT layers to refine visual cues during each iteration and does not require additional target speaker labels. The improved experimental results demonstrate that employing an iterative cue encoder, which integrates AV-HuBERT layers, yields more robust audio-visual correspondence compared to solely utilizing viseme-phoneme visual embeddings.\\\indent It is worth noting that after utilizing the MAR strategy, the SI-SDR can be further improved from 11.728 to 12.331 while PESQ improved from 2.765 to 2.922, and STOI improved from 0.878 to 0.887. Both the intelligible and comprehensible improvements could indicate that the final estimated target speech embedding becomes much clearer and more accurate after undergoing refinement using the proposed MAR strategy. It could also verify the rationality of the designed loss function, which jointly optimized three tasks, including masked region prediction, unmasked region prediction, and final target speech signal prediction. 
\subsection{Effect of Different Mask Duration for MAR Strategy}
\begin{table}[t]
\begin{center}
\caption{The first column is different mask duration gaps in the mixture speech waveform, the SI-SDR, and SI-SDRi are reported with AVHuMAR-TSE System.} \label{GAP}
\begin{tabular}{p{0.15\textwidth}p{0.1\textwidth}p{0.1\textwidth}} 
  \toprule
  \textbf{Mask Duration (ms)}& \textbf{SI-SDR $(\uparrow)$}& \textbf{SI-SDRi $(\uparrow)$} \\
 \toprule
  \centering 100&12.292 & 12.338\\
  \centering 200&11.956&12.012\\
  \centering 300&\textbf{12.331}&\textbf{12.379}\\
  \centering 400&11.925&11.973\\
  \centering 500&11.826&11.873\\
  \centering 600&11.695&11.742\\
 \bottomrule
\end{tabular}
\end{center}
\end{table}
To investigate the effects of varying mask duration gaps on the final target speech extraction performance of the proposed AVHuMAR-TSE system, we present the SI-SDR and SI-SDRi values as the mask duration gaps increase from 100 ms to 600 ms, using 100 ms intervals. As shown in TABLE \ref{GAP}, when the mask duration gap is set to 300 ms, AVHuMAR-TSE could achieve the best performances in both SI-SDR and SI-SDRi. Furthermore, the results with mask duration gaps of 100 ms and 200 ms exhibit marginally better performances compared to those with mask duration gaps of 400 ms and above.
Except for the 600 ms mask duration gap, both SI-SDR and SI-SDRi results are higher than those obtained without employing the MAR strategy. Such experimental results are reasonable, as the total duration of the input mixture waveform is 4 seconds, and recovering a mask duration gap of 600 ms is relatively challenging. More specifically, the MAR strategy aims at learning speech context features via explicit mask-recover embedding loss $L_{RECOVER}$ and TSE embedding loss $L_{TSE\_embedding}$. With guidance from both intermediate estimated target speech context and the target speaker's lip movements, the boundary of the masked region could be relatively easy to recover. However, the center area of the masked region could still be hard to recover even with guidance from both modalities. Based on this analysis, too much mask gap duration could not be conducive to the MAR strategy. On the contrary, it may even bring some adverse effects. The model may learn some corrupted audio-visual correlations and the overall model weights might be biased towards the recovery task and forget the TSE knowledge learned in the first training stage.
  
\subsection{Case Study}
To explore the performance improvement details and show the effectiveness of the MAR strategy more intuitively, we visualize the extracted target speech spectrograms.
\begin{figure}[!h]
\centering
 \centering
  \includegraphics[scale=0.32]{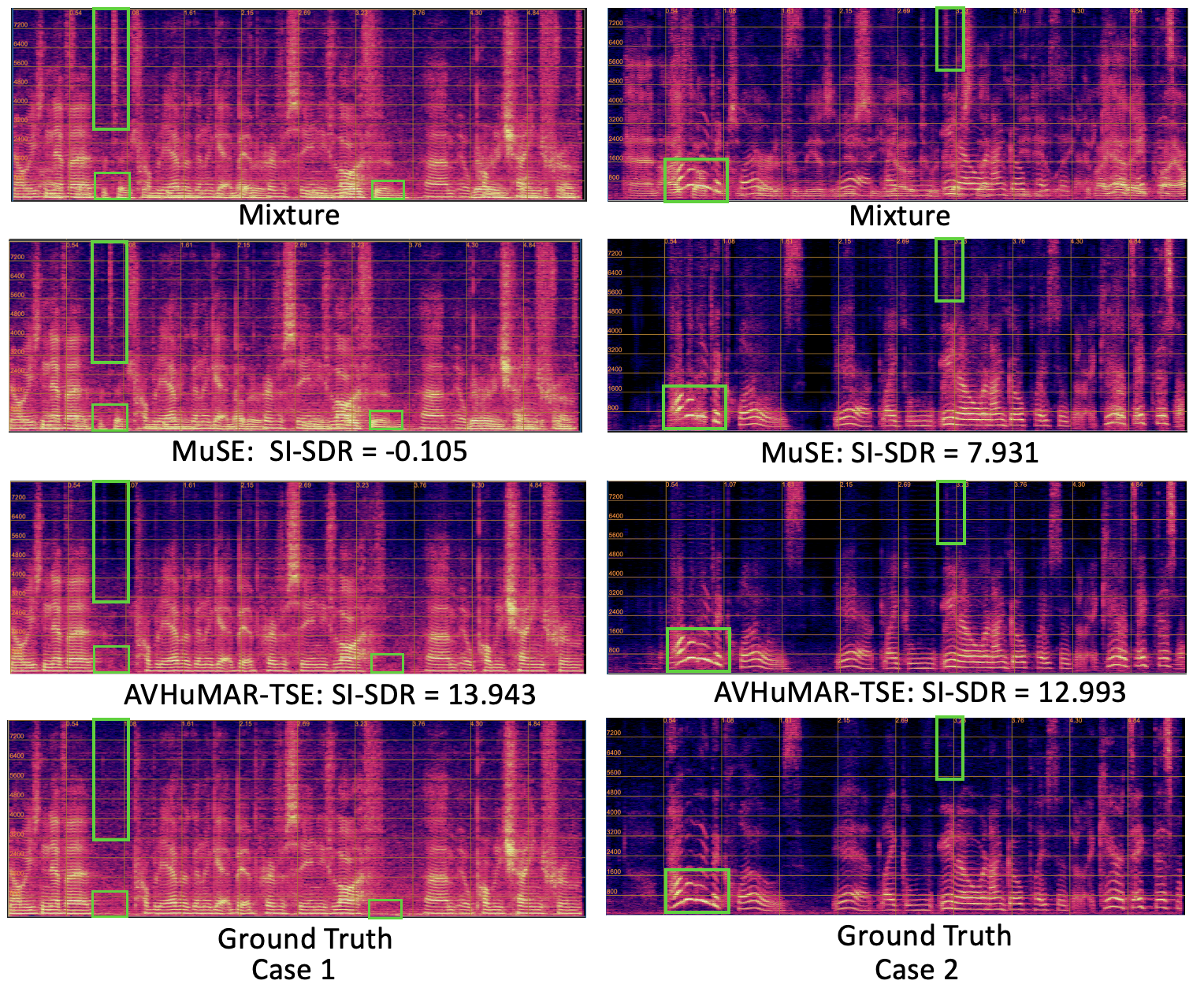}
   \caption{Comparison of target speech spectrograms extracted by AVHuMAR-TSE system and MuSE system.}
    \label{compare}
\end{figure}
As shown in Fig. \ref{compare}, we visualize two cases. For each column, the spectrograms from top to down are raw mixture speech, the target speech extracted by the best baseline MuSE system, the target speech extracted by our proposed AVHuMAR-TSE system, and the target speech ground truth. For case 1, AVHuMAR-TSE could achieve 13.943 on SI-SDR while MuSE could only achieve -0.105. We marked three obvious extraction failed regions with green boxes in the second spectrogram from MuSE. As we can observe, in the top-left box, MuSE incorrectly extracted high-frequency voice segments from another speaker, while the two bottom boxes show that MuSE mistakenly extracted low-frequency parts belonging to another speaker. Compared to MuSE, the third spectrogram obtained from AVHuMAR-TSE appears to be very close to the ground truth. For case 2, MuSE achieves SI-SDR with 7.931 while the AVHuAMR-TSE system could achieve 12.993. In this case, MuSE manages to extract coarse target speech but still misses some low-frequency parts for target speech, which is marked in the left-bottom box. Besides, MuSE also falsely extracted some high-frequency parts from another speaker, which is marked in the right-top box. 
 
\section{Conclusion}
In this work, we integrate pre-trained AV-HuBERT layers into an AV-TSE system, which is called the AVHuBERT-TSE system. To further enhance both intra and inter-modality alignments, thereby improving the overall extraction performance, we propose the AVHuMAR-TSE system. Compared to the three time-domain AV-TSE baseline systems, the proposed AVHuMAR-TSE system shows significant performance improvements in terms of both subjective and objective metrics on the VoxCeleb2 dataset. Such results prove that both the pre-trained AV-HuBERT layers and the proposed MAR strategy could enhance audio-visual correspondence and speech context correlation. Experimental results with different mask duration gaps show that it is critical to select an appropriate mask for the input mixture speech waveform. In the future, we plan to delve deeper into how the MAR strategy can align the audio-visual latent feature space and expand our AVHuMAR-TSE system to different mixture scenarios.
 
\bibliographystyle{IEEEtran}
\bibliography{refs} 

\begin{thebibliography}{10}
\providecommand{\url}[1]{#1}
\csname url@samestyle\endcsname
\providecommand{\newblock}{\relax}
\providecommand{\bibinfo}[2]{#2}
\providecommand{\BIBentrySTDinterwordspacing}{\spaceskip=0pt\relax}
\providecommand{\BIBentryALTinterwordstretchfactor}{4}
\providecommand{\BIBentryALTinterwordspacing}{\spaceskip=\fontdimen2\font plus
\BIBentryALTinterwordstretchfactor\fontdimen3\font minus \fontdimen4\font\relax}
\providecommand{\BIBforeignlanguage}[2]{{%
\expandafter\ifx\csname l@#1\endcsname\relax
\typeout{** WARNING: IEEEtran.bst: No hyphenation pattern has been}%
\typeout{** loaded for the language `#1'. Using the pattern for}%
\typeout{** the default language instead.}%
\else
\language=\csname l@#1\endcsname
\fi
#2}}
\providecommand{\BIBdecl}{\relax}
\BIBdecl

\bibitem{token2vec}
X.~Yue, J.~Ao, X.~Gao, and H.~Li, ``Token2vec: A joint self-supervised pre-training framework using unpaired speech and text,'' in \emph{ICASSP 2023 - 2023 IEEE International Conference on Acoustics, Speech and Signal Processing (ICASSP)}, 2023, pp. 1--5.

\bibitem{spkverify}
R.~Tao, K.~A. Lee, Z.~Shi, and H.~Li, ``Speaker recognition with two-step multi-modal deep cleansing,'' in \emph{ICASSP 2023 - 2023 IEEE International Conference on Acoustics, Speech and Signal Processing (ICASSP)}, 2023, pp. 1--5.

\bibitem{speech-translation}
S.~Ouyang, R.~Ye, and L.~Li, ``{WACO}: Word-aligned contrastive learning for speech translation,'' in \emph{Proceedings of the 61st Annual Meeting of the Association for Computational Linguistics (Volume 1: Long Papers)}.\hskip 1em plus 0.5em minus 0.4em\relax Association for Computational Linguistics, Jul. 2023, pp. 3891--3907.

\bibitem{chen2022hilvoice}
X.~Chen, Q.~Huang, X.~Wu, Z.~Wu, and H.~Meng, ``Hilvoice: Human-in-the-loop style selection for elder-facing speech synthesis,'' in \emph{2022 13th International Symposium on Chinese Spoken Language Processing (ISCSLP)}.\hskip 1em plus 0.5em minus 0.4em\relax IEEE, 2022, pp. 86--90.

\bibitem{brain}
N.~Mesgarani, ``Robust speech processing in human auditory cortex,'' \emph{The Journal of the Acoustical Society of America}, vol. 143, no.~3, pp. 1744--1744, 2018.

\bibitem{upit}
M.~Kolbæk, D.~Yu, Z.-H. Tan, and J.~Jensen, ``Multitalker speech separation with utterance-level permutation invariant training of deep recurrent neural networks,'' \emph{IEEE/ACM Transactions on Audio, Speech, and Language Processing}, vol.~25, no.~10, pp. 1901--1913, 2017.

\bibitem{Conv-TasNet}
Y.~Luo and N.~Mesgarani, ``Conv-tasnet: Surpassing ideal time–frequency magnitude masking for speech separation,'' \emph{IEEE/ACM Transactions on Audio, Speech, and Language Processing}, vol.~27, no.~8, pp. 1256--1266, 2019.

\bibitem{TSE-review}
K.~Zmolikova, M.~Delcroix, T.~Ochiai, K.~Kinoshita, J.~Černocký, and D.~Yu, ``Neural target speech extraction: An overview,'' \emph{IEEE Signal Processing Magazine}, vol.~40, no.~3, pp. 8--29, 2023.

\bibitem{select-listen}
Z.~Pan, R.~Tao, C.~Xu, and H.~Li, ``Selective listening by synchronizing speech with lips,'' \emph{IEEE/ACM Transactions on Audio, Speech, and Language Processing}, vol.~30, pp. 1--1, 01 2022.

\bibitem{SpeakerBeam}
K.~Žmolíková, M.~Delcroix, K.~Kinoshita, T.~Ochiai, T.~Nakatani, L.~Burget, and J.~Černocký, ``Speakerbeam: Speaker aware neural network for target speaker extraction in speech mixtures,'' \emph{IEEE Journal of Selected Topics in Signal Processing}, vol.~13, no.~4, pp. 800--814, 2019.

\bibitem{VoiceFilter}
Q.~Wang, H.~Muckenhirn, K.~W. Wilson, P.~Sridhar, Z.~Wu, J.~R. Hershey, R.~A. Saurous, R.~J. Weiss, Y.~Jia, and I.~L{\'o}pez-Moreno, ``Voicefilter: Targeted voice separation by speaker-conditioned spectrogram masking,'' in \emph{Interspeech}, 2018.

\bibitem{SpEx}
C.~Xu, W.~Rao, E.~S. Chng, and H.~Li, ``Spex: Multi-scale time domain speaker extraction network,'' \emph{IEEE/ACM Transactions on Audio, Speech, and Language Processing}, vol.~28, pp. 1370--1384, 2020.

\bibitem{SpEx+}
M.~Ge, C.~Xu, L.~Wang, C.~E. Siong, J.~Dang, and H.~Li, ``Spex+: A complete time domain speaker extraction network,'' in \emph{Interspeech}, 2020.

\bibitem{Rethinking}
J.~Li, M.~Ge, Z.~Pan, R.~Cao, L.~Wang, J.~Dang, and S.~Zhang, ``{Rethinking the Visual Cues in Audio-Visual Speaker Extraction},'' in \emph{Proc. INTERSPEECH 2023}, 2023, pp. 3754--3758.

\bibitem{ImagineNET}
Z.~Pan, W.~Wang, M.~Borsdorf, and H.~Li, ``Imaginenet: Target speaker extraction with intermittent visual cue through embedding inpainting,'' in \emph{ICASSP 2023 - 2023 IEEE International Conference on Acoustics, Speech and Signal Processing (ICASSP)}, 2023, pp. 1--5.

\bibitem{VisualVoice}
R.~Gao and K.~Grauman, ``Visualvoice: Audio-visual speech separation with cross-modal consistency,'' \emph{2021 IEEE/CVF Conference on Computer Vision and Pattern Recognition (CVPR)}, pp. 15\,490--15\,500, 2021.

\bibitem{ssl_speech_review}
A.~rahman Mohamed, H.~yi~Lee, L.~Borgholt, J.~D. Havtorn, J.~Edin, C.~Igel, K.~Kirchhoff, S.-W. Li, K.~Livescu, L.~Maal{\o}e, T.~N. Sainath, and S.~Watanabe, ``Self-supervised speech representation learning: A review,'' \emph{IEEE Journal of Selected Topics in Signal Processing}, vol.~16, pp. 1179--1210, 2022.

\bibitem{chen2022character}
X.~Chen, C.~Song, Y.~Zhou, Z.~Wu, C.~Chen, Z.~Wu, and H.~Meng, ``A character-level span-based model for mandarin prosodic structure prediction,'' in \emph{ICASSP 2022-2022 IEEE International Conference on Acoustics, Speech and Signal Processing (ICASSP)}.\hskip 1em plus 0.5em minus 0.4em\relax IEEE, 2022, pp. 7602--7606.

\bibitem{chen2022unsupervised}
X.~Chen, S.~Lei, Z.~Wu, D.~Xu, W.~Zhao, and H.~Meng, ``Unsupervised multi-scale expressive speaking style modeling with hierarchical context information for audiobook speech synthesis,'' in \emph{Proceedings of the 29th International Conference on Computational Linguistics}, 2022, pp. 7193--7202.

\bibitem{chen2023stylespeech}
X.~Chen, X.~Wang, S.~Zhang, L.~He, Z.~Wu, X.~Wu, and H.~Meng, ``Stylespeech: Self-supervised style enhancing with vq-vae-based pre-training for expressive audiobook speech synthesis,'' \emph{arXiv preprint arXiv:2312.12181}, 2023.

\bibitem{wav2vec2}
A.~Baevski, Y.~Zhou, A.~Mohamed, and M.~Auli, ``wav2vec 2.0: A framework for self-supervised learning of speech representations,'' in \emph{Advances in Neural Information Processing Systems}, vol.~33.\hskip 1em plus 0.5em minus 0.4em\relax Curran Associates, Inc., 2020, pp. 12\,449--12\,460.

\bibitem{HuBERT}
W.-N. Hsu, B.~Bolte, Y.-H. Tsai, K.~Lakhotia, R.~Salakhutdinov, and A.~Mohamed, ``Hubert: Self-supervised speech representation learning by masked prediction of hidden units,'' \emph{IEEE/ACM Transactions on Audio, Speech, and Language Processing}, vol.~PP, pp. 1--1, 10 2021.

\bibitem{Huang2022InvestigatingSL}
Z.~Huang, S.~Watanabe, S.~wen Yang, L.~P. Garc{\'i}a-Perera, and S.~Khudanpur, ``Investigating self-supervised learning for speech enhancement and separation,'' \emph{ICASSP 2022 - 2022 IEEE International Conference on Acoustics, Speech and Signal Processing (ICASSP)}, pp. 6837--6841, 2022.

\bibitem{WavLM}
S.~Chen, C.~Wang, Z.~Chen, Y.~Wu, S.~Liu, Z.~Chen, J.~Li, N.~Kanda, T.~Yoshioka, X.~Xiao, J.~Wu, L.~Zhou, S.~Ren, Y.~Qian, Y.~Qian, M.~Zeng, and F.~Wei, ``Wavlm: Large-scale self-supervised pre-training for full stack speech processing,'' \emph{IEEE Journal of Selected Topics in Signal Processing}, vol.~16, pp. 1505--1518, 2021.

\bibitem{wavlm-ss}
Z.~Chen, N.~Kanda, J.~Wu, Y.~Wu, X.~Wang, T.~Yoshioka, J.~Li, S.~Sivasankaran, and S.~E. Eskimez, ``Speech separation with large-scale self-supervised learning,'' in \emph{ICASSP 2023 - 2023 IEEE International Conference on Acoustics, Speech and Signal Processing (ICASSP)}, 2023, pp. 1--5.

\bibitem{chen2024exploiting}
X.~Chen, Y.~Wang, X.~Wu, D.~Wang, Z.~Wu, X.~Liu, and H.~Meng, ``Exploiting audio-visual features with pretrained av-hubert for multi-modal dysarthric speech reconstruction,'' in \emph{ICASSP 2024-2024 IEEE International Conference on Acoustics, Speech and Signal Processing (ICASSP)}.\hskip 1em plus 0.5em minus 0.4em\relax IEEE, 2024.

\bibitem{AVhubert}
B.~Shi, W.-N. Hsu, K.~Lakhotia, and A.~Mohamed, ``Learning audio-visual speech representation by masked multimodal cluster prediction,'' in \emph{International Conference on Learning Representations}, 2022.

\bibitem{muse}
Z.~Pan, R.~Tao, C.~Xu, and H.~Li, ``Muse: Multi-modal target speaker extraction with visual cues,'' in \emph{ICASSP 2021 - 2021 IEEE International Conference on Acoustics, Speech and Signal Processing (ICASSP)}, 2021, pp. 6678--6682.

\bibitem{tdse}
J.~Wu, Y.~Xu, S.-X. Zhang, L.-W. Chen, M.~Yu, L.~Xie, and D.~Yu, ``Time domain audio visual speech separation,'' in \emph{2019 IEEE Automatic Speech Recognition and Understanding Workshop (ASRU)}, 2019, pp. 667--673.

\bibitem{usev}
Z.~Pan, M.~Ge, and H.~Li, ``Usev: Universal speaker extraction with visual cue,'' \emph{IEEE/ACM Trans. Audio, Speech and Lang. Proc.}, vol.~30, p. 3032–3045, sep 2022.

\bibitem{adapter}
S.~Otake, R.~Kawakami, and N.~Inoue, ``Parameter efficient transfer learning for various speech processing tasks,'' in \emph{ICASSP 2023 - 2023 IEEE International Conference on Acoustics, Speech and Signal Processing (ICASSP)}, 2023, pp. 1--5.

\bibitem{sidecar}
L.~Meng, J.~Kang, M.~Cui, Y.~Wang, X.~Wu, and H.~Meng, ``A sidecar separator can convert a single-talker speech recognition system to a multi-talker one,'' in \emph{ICASSP 2023 - 2023 IEEE International Conference on Acoustics, Speech and Signal Processing (ICASSP)}, 2023, pp. 1--5.

\bibitem{still}
T.~Afouras, J.~S. Chung, and A.~Zisserman, ``My lips are concealed: Audio-visual speech enhancement through obstructions,'' in \emph{Interspeech}, 2019.

\bibitem{still-image}
S.~Chung, S.~Choe, J.~S. Chung, and H.~Kang, ``Facefilter: Audio-visual speech separation using still images,'' in \emph{Interspeech 2020}.\hskip 1em plus 0.5em minus 0.4em\relax {ISCA}, pp. 3481--3485.

\bibitem{speech-inpainting}
J.~F. Montesinos, D.~Michelsanti, G.~Haro, Z.-H. Tan, and J.~Jensen, ``{Speech inpainting: Context-based speech synthesis guided by video},'' in \emph{Proc. INTERSPEECH 2023}, 2023, pp. 4459--4463.

\bibitem{Phoneme-to-viseme}
H.~Bear and R.~Harvey, ``Phoneme-to-viseme mappings: The good, the bad, and the ugly,'' \emph{Speech Communication}, 07 2017.

\bibitem{MEAN-SQUARED-ERROR}
K.~Das, J.~Jiang, and J.~Rao, ``Mean squared error of empirical predictor,'' 2004.

\bibitem{voxceleb2}
J.~S. Chung, A.~Nagrani, and A.~Zisserman, ``Voxceleb2: Deep speaker recognition,'' in \emph{INTERSPEECH}, 2018.

\bibitem{SDR}
J.~L. Roux, S.~Wisdom, H.~Erdogan, and J.~R. Hershey, ``Sdr – half-baked or well done?'' \emph{ICASSP 2019 - 2019 IEEE International Conference on Acoustics, Speech and Signal Processing (ICASSP)}, pp. 626--630, 2018.

\bibitem{pesq}
S.~{Basterrech}, G.~{Rubino}, and M.~{Varela}, ``{Single-sided Real-time PESQ Score Estimation},'' \emph{arXiv e-prints}, p. arXiv:1212.6350, Dec. 2012.

\bibitem{stoi}
C.~H. Taal, R.~C. Hendriks, R.~Heusdens, and J.~R. Jensen, ``An algorithm for intelligibility prediction of time–frequency weighted noisy speech,'' \emph{IEEE Transactions on Audio, Speech, and Language Processing}, vol.~19, pp. 2125--2136, 2011.

\end{thebibliography}
\end{document}